\documentclass[reprint,
 amsmath,amssymb,
 aps,
]{revtex4-2}

\usepackage{graphicx}
\usepackage{dcolumn}
\usepackage{bm}
\usepackage{xcolor}

\begin{document}


\title{Canceling and inverting normal and anomalous group-velocity dispersion using space-time wave packets}

\author{Layton A. Hall$^{1}$}
\author{Ayman F. Abouraddy$^{1,*}$}
\affiliation{$^{1}$CREOL, The College of Optics \& Photonics, University of Central~Florida, Orlando, FL 32816, USA}
\affiliation{$^*$Corresponding author: raddy@creol.ucf.edu}

\begin{abstract}
Angular dispersion can counterbalance normal group-velocity dispersion (GVD) that increases the wave-vector length in a dispersive medium. By tilting the wave vector, angular dispersion reduces the axial wave number in this case to match the pre-GVD value. By the same token, however, angular dispersion fails to counterbalance anomalous GVD, which in contrast reduces the wave-vector length. Consequently, GVD-cancellation via angular dispersion has not been demonstrated to date in the anomalous dispersion regime. We synthesize here structured femtosecond pulsed beams known as `space-time' wave packets designed to realize dispersion-cancellation symmetrically in either the normal- or anomalous-GVD regimes by virtue of non-differentiable angular dispersion inculcated into the pulsed field. Furthermore, we also verify GVD-inversion: reversing the GVD sign experienced by the field with respect to that dictated by the chromatic dispersion of the medium itself. 
\end{abstract}


\maketitle

\section{Introduction}

Chromatic dispersion resulting from the wavelength-dependence of the refractive index is an inescapable feature of optical materials, which leads to pulse broadening and distortion \cite{SalehBook07,Weiner09Book}. One may combat its impact via dispersion \textit{compensation} or dispersion \textit{cancellation}. In the former, dispersive broadening and the associated chirp are compensated after (or pre-compensated before) passage through the medium, which is key to the success of chirped pulsed amplification (CPA) \cite{Strickland85OC}, for example. Normal group-velocity dispersion (GVD) can be compensated by a pair of gratings or prisms \cite{Fork84OL}, anomalous GVD by a Martinez stretcher \cite{Martinez87IEEEJQE}, and almost arbitrary dispersion by a $4f$ pulse shaper or other techniques \cite{White93OL,Lemoff93OL,Kane97JOSAB,Kane97JOSAB2,Weiner00RSI,Runge20NP}. In all these cases, it is the group delay dispersion (GDD) that is being neutralized. More challenging, however, is to neutralize dispersion \textit{during} passage through a dispersive medium, so that the pulse travels invariantly, which we refer to as \textit{dispersion cancellation}. Such a capability is crucial, for instance, in enabling efficient nonlinear interactions in long crystals. Angular dispersion \cite{Torres10AOP}, whereby each frequency in the pulse is directed at a prescribed angle, has been successfully utilized for this purpose \cite{Szatmari96OL}. The resulting field structure after inculcating angular dispersion is typically known as a tilted pulse front (TPF) \cite{Fulop10Review}. To date, however, angular dispersion has been used for dispersion cancellation in only the \textit{normal}-GVD regime. There have been no reports of dispersion-cancellation in the anomalous-GVD regime, and well-established theoretical considerations suggest the impossibility of such a goal \cite{Martinez84JOSAA}. Indeed, there is strong \textit{a priori} conceptual support for such a claim. Because normal GVD \textit{increases} the wave-vector length in the dispersive medium by a frequency-dependent amount, Changing the wave-vector angle for each frequency can reduce its axial component to the pre-GVD value. Anomalous GVD, on the other hand, \textit{reduces} the wave-vector length, and no angular tilt can compensate for such a reduction. Consequently, cancelling anomalous GVD has \textit{not} been reported to date.

Nevertheless, pulsed beams or wave packets that are propagation invariant (diffraction-free and dispersion-free) in dispersive media by virtue of their spatio-temporal field structure have been known to exist theoretically \cite{Porras03PRE2,Porras03OL,Longhi04OL,Porras04PRE,Christodoulides04OL,Mills12PRA} in presence of normal \textit{and} \cite{Malaguti09PRA} anomalous \cite{Malaguti08OL} GVD. These sought-after dispersion-free structured fields have \textit{not} been observed to date in linear dispersive media, although there is evidence for their presence in nonlinear interactions \cite{DiTrapani03PRL,Faccio06PRL,Faccio07OE,Porras07JOSAB}. Such spatiotemporally structured fields have been recently studied systematically in free space and non-dispersive dielectrics under the rubric of `space-time' (ST) wave packets \cite{Kondakci16OE,Parker16OE,Porras17OL,Efremidis17OL,PorrasPRA18,Yessenov19OPN,Yessenov22AOP}. These propagation-invariant wave packets feature a unique set of characteristics including tunable group velocities in absence of chromatic dispersion \cite{Wong17ACSP2,Kondakci19NC,Yessenov19OE,Yessenov19OL}, self-healing \cite{Kondakci18OL}, anomalous refraction \cite{Bhaduri20NP,Motz21OL,Yessenon21JOSAA1,Motz21JOSAA,Yessenon21JOSAA2}, and novel ST Talbot effects \cite{Hall21APLP}. The central characteristics of ST wave packets in free space are now well-understood \cite{Kondakci19OL,Yessenov22AOP}, and potential applications are being evaluated in optical communications \cite{Bhaduri19OL,Yessenov20NC} and device physics \cite{Shiri20NC,Schepler20ACSP,Shiri20OL,Kibler2021PRL,Bejot2021ACSP,Ruano21JO,Guo21PRR}.

In contrast to TPFs, the angular dispersion undergirding ST wave packets is `non-differentiable': the derivative of the propagation angle is \textit{not} defined at one wavelength \cite{Hall21OL,Hall21OL3NormalGVD,Hall22OEConsequences}. Such angular dispersion can be produced by a recently developed pulsed-beam shaper that serves as a universal angular-dispersion synthesizer in one dimension \cite{Hall21OEUniversal}. We have recently shown that non-differentiable angular dispersion is the crucial ingredient for tuning the group velocity of a ST wave packet and introducing an arbitrary dispersion profile in free space \cite{Yessenov21ACSP,Hall21OL3NormalGVD}. This suggests the prospect for dispersion-cancellation in presence of either normal- or anomalous-GVD. 

Here, we verify the propagation invariance of ST wave packets symmetrically in \textit{both} the normal- \textit{and} anomalous-GVD regimes, the latter for the first time to the best of our knowledge. We identify the spatio-temporal spectral structures needed for achieving GVD-cancellation. The crucial step is to first change the ST wave-packet group velocity via non-differentiable angular dispersion, which opens up a space for subsequent decrease or increase in the axial wave vector relative to the dispersion-free configuration. We sculpt the spatio-temporal spectrum of $\approx\!200$-fs ($\approx\!16$-nm-bandwidth) pulses to produce ST wave packets that are propagation invariant at a wavelength of $\approx\!1$~$\mu$m in ZnSe as a representative normal-GVD medium, and chirped Bragg mirrors that produce anomalous GVD. Uniquely, the GVD in the medium is cancelled while maintaining independent control over the group velocity of the ST wave packet (in both the subluminal and superluminal regimes). Furthermore, not only is propagation invariance achieved in dispersive media, but dispersion inversion is also verified: the sign of GVD experienced by the wave packet in the medium is \textit{reversed}. We thus produce wave packets that undergo normal GVD in an anomalously dispersive medium, and vice versa. We expect these results to be particularly useful in tailoring multi-wavelength nonlinear optical interactions in long crystals.

\begin{figure}[t!]
\centering
\includegraphics[width=8.6cm]{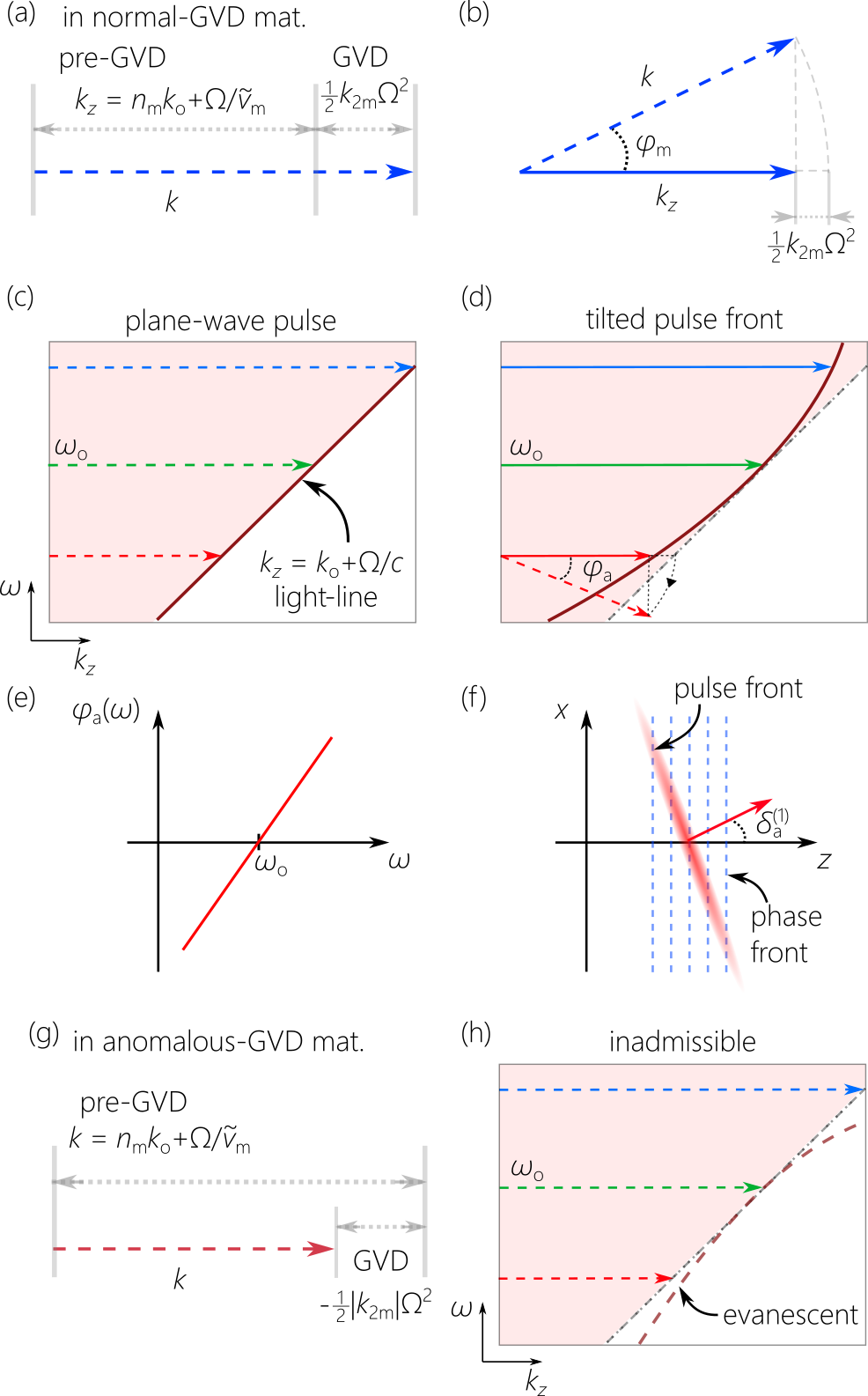}
\caption{(a) Normal GVD \textit{increases} the wave-vector length, (b) but angular dispersion reduces the axial wave-vector component $k_{z}$ to the pre-GVD value. (c) In free space, $k(\omega)\!=\!\tfrac{\omega}{c}\!=\!k_{\mathrm{o}}+\tfrac{\Omega}{c}$, and (d) after rotating $k(\omega)$ by an angle $\varphi_{\mathrm{a}}(\omega)$, the field experiences anomalous GVD in free space. (e) The requisite angle introduced into the plane-wave pulse in free space is $\varphi_{\mathrm{a}}(\omega)\!=\!\varphi_{\mathrm{a}}(\omega_{\mathrm{o}}+\Omega)\!\propto\!\Omega$. (f) The angular dispersion in (e) produces a field in the form of a tilted pulse front (TPF). (g) In the \textit{anomalous}-GVD regime, the angular-dispersion approach fails because the wave-vector length is \textit{reduced}, and a tilt can\textit{not} increase $k_{z}$ to its pre-GVD value. (h) Introducing normal GVD into a plane-wave pulse in free space via conventional angular dispersion renders the field evanescent.}
\label{Fig:DiagramFoDifferentiable}
\end{figure}

\begin{figure*}[t!]
\centering
\includegraphics[width=16cm]{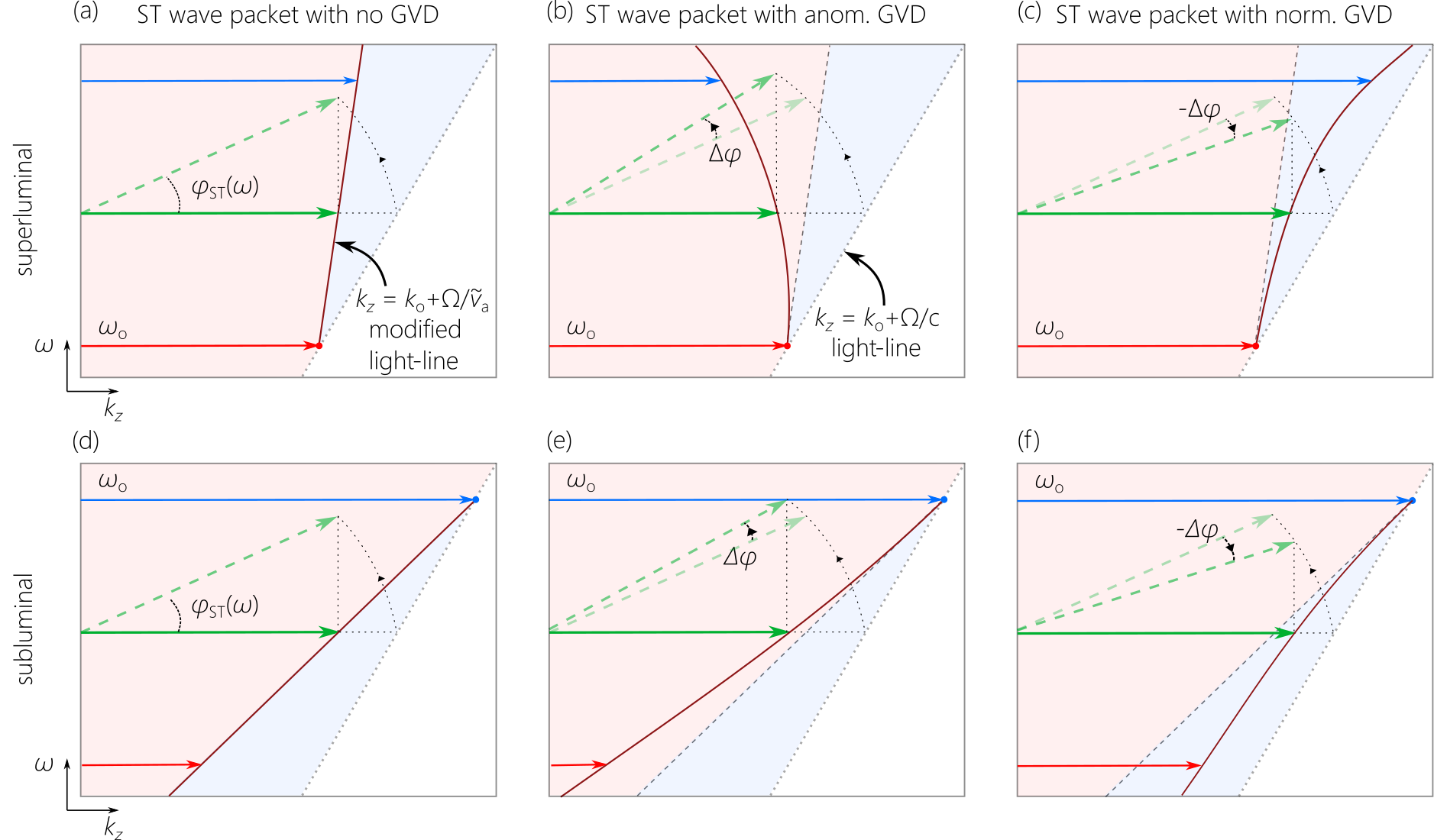}
\caption{The impact of \textit{non-differentiable} angular dispersion on the group velocity and GVD in free space. (a) By introducing a tilt angle $\varphi_{\mathrm{ST}}(\omega_{\mathrm{o}}+\Omega)\!\approx\!\eta\sqrt{\Omega/\omega_{\mathrm{o}}}$ into a plane-wave pulse, the group velocity becomes $\widetilde{v}\!=\!c/\widetilde{n}_{\mathrm{a}}$, and $k_{z}$ is determined by a modified light-line $k_{z}\!=\!k_{\mathrm{o}}+\Omega/\widetilde{v}_{\mathrm{a}}$ rather than the free-space light-line $k_{z}\!=\!\tfrac{\omega}{c}\!=\!k_{\mathrm{o}}+\Omega/c$. (b) By further increasing the tilt angle to $\varphi_{\mathrm{ST}}+\Delta\varphi$, $k_{z}$ is reduced and determined by a quadratic curve consistent with anomalous GVD in free space. (c) By reducing the tilt angle to $\varphi_{\mathrm{ST}}-\Delta\varphi$, $k_{z}$ is increased and determined by a quadratic curve consistent with normal GVD. Such scenarios can be achieved only via non-differentiable angular dispersion. Panels (a-c) correspond to superluminal and (d-f) to subluminal ST wave packets.}
\label{Fig:DiagramForNonDifferentiable}
\end{figure*}
 
\section{The challenge of canceling anomalous GVD via angular dispersion}

Angular dispersion, whereby each frequency $\omega$ in a pulsed field travels at a different angle $\varphi(\omega)$ \cite{Torres10AOP}, produces anomalous GVD in free space \cite{Martinez84JOSAA,Fork84OL,Gordon84OL}, which can help cancel normal GVD \cite{Szatmari96OL,Porras03PRE2}. In fact, it is commonly understood that angular dispersion can\textit{not} produce normal GVD in free space \cite{Martinez84JOSAA}, and thus does not provide the possibility of cancelling anomalous GVD. The illustration in Fig.~\ref{Fig:DiagramFoDifferentiable} elucidates the origin of this asymmetry between realizing normal and anomalous GVD via conventional angular dispersion.

Consider a plane-wave pulse traveling in a dispersive medium of refractive index $n(\omega)$, and expand the wave number $k(\omega)\!=\!n(\omega)\omega/c$ around a frequency $\omega_{\mathrm{o}}$, $k(\omega_{\mathrm{o}}+\Omega)\!\approx\!n_{\mathrm{m}}k_{\mathrm{o}}+\tfrac{\Omega}{\widetilde{v}_{\mathrm{m}}}+\tfrac{1}{2}k_{2\mathrm{m}}\Omega^{2}$; here $\Omega\!=\!\omega-\omega_{\mathrm{o}}$, $k_{\mathrm{o}}\!=\!\omega_{\mathrm{o}}/c$, $c$ is the speed of light in vacuum, $n_{\mathrm{m}}\!=\!n(\omega_{\mathrm{o}})$, $\widetilde{v}_{\mathrm{m}}\!=\!1/\tfrac{dk}{d\omega}\big|_{\omega_{\mathrm{o}}}\!=\!\tfrac{c}{\widetilde{n}_{\mathrm{m}}}$ is the group velocity and $\widetilde{n}_{\mathrm{m}}$ the group index, and $k_{2\mathrm{m}}\!=\!\tfrac{d^{2}k}{d\omega^{2}}\big|_{\omega_{\mathrm{o}}}$ is the GVD coefficient \cite{SalehBook07}. Throughout, we use the subscript `m' to denote quantities in the dispersive medium, and the subscript `a' for the corresponding quantities in free space. It is clear that normal GVD $k_{2\mathrm{m}}\!>\!0$ \textit{increases} the wave-vector length by a frequency-dependent amount $\tfrac{1}{2}k_{2\mathrm{m}}\Omega^{2}$ [Fig.~\ref{Fig:DiagramFoDifferentiable}(a)], which can be counterbalanced by tilting the wave vector by an angle $\varphi_{\mathrm{m}}(\omega)$ to \textit{reduce} the axial component of the wave vector in the medium to the pre-GVD value $k_{z}(\omega)\!=\!k(\omega)\cos{\{\varphi_{\mathrm{m}}(\omega)\}}\!\approx\!n_{\mathrm{m}}k_{\mathrm{o}}+\tfrac{\Omega}{\widetilde{v}_{\mathrm{m}}}$ [Fig.~\ref{Fig:DiagramFoDifferentiable}(b)]. To realize this condition in the small-angle limit starting with a plane-wave pulse in free space [Fig.~\ref{Fig:DiagramFoDifferentiable}(c)], we tilt the wave vector associated with $\omega$ by an angle $\varphi_{\mathrm{a}}(\Omega)\!\approx\!\tfrac{\Omega}{\omega_{\mathrm{o}}}\tan{\delta_{\mathrm{a}}^{(1)}}$ [Fig.~\ref{Fig:DiagramFoDifferentiable}(d,e)], where $\delta_{\mathrm{a}}^{(1)}$ is the tilt angle of the pulse front (the plane of constant amplitude) with respect to the phase front (the plane of constant phase) [Fig.~\ref{Fig:DiagramFoDifferentiable}(f)]. Such a field structure is known as a TPF \cite{Fulop10Review}, which in general experiences \textit{anomalous} GVD in free space  $k_{2\mathrm{a}}\!=\!-\tfrac{1}{c\omega_{\mathrm{o}}}\tan^{2}{\delta_{\mathrm{o}}^{(1)}}$ \cite{Porras03PRE2}. Once coupled to a medium in its normal-GVD regime, the TPF propagates dispersion-free if $k_{2\mathrm{a}}\!=\!-n_{\mathrm{m}}k_{2\mathrm{m}}$ (Appendix).

The challenge of cancelling \textit{anomalous} GVD $k_{2\mathrm{m}}\!<\!0$ is now clear. Anomalous GVD \textit{reduces} $k(\omega)$ in the medium by $-\tfrac{1}{2}|k_{2\mathrm{m}}|\Omega^{2}$ [Fig.~\ref{Fig:DiagramFoDifferentiable}(g)], and no angular tilt can \textit{increase} $k_{z}$ to the pre-GVD value. As shown in Fig.~\ref{Fig:DiagramFoDifferentiable}(h), the dispersion curve for normal GVD in free space lies \textit{below} the light-line, corresponding to an evanescent field. Although this appears to be an insurmountable obstacle, we show here that it is resolved by ST wave packets endowed with \textit{non-differentiable} angular dispersion.

\section{Theory of propagation-invariant space-time wave packets in dispersive media}

\begin{figure*}[t!]
\centering
\includegraphics[width=17.6cm]{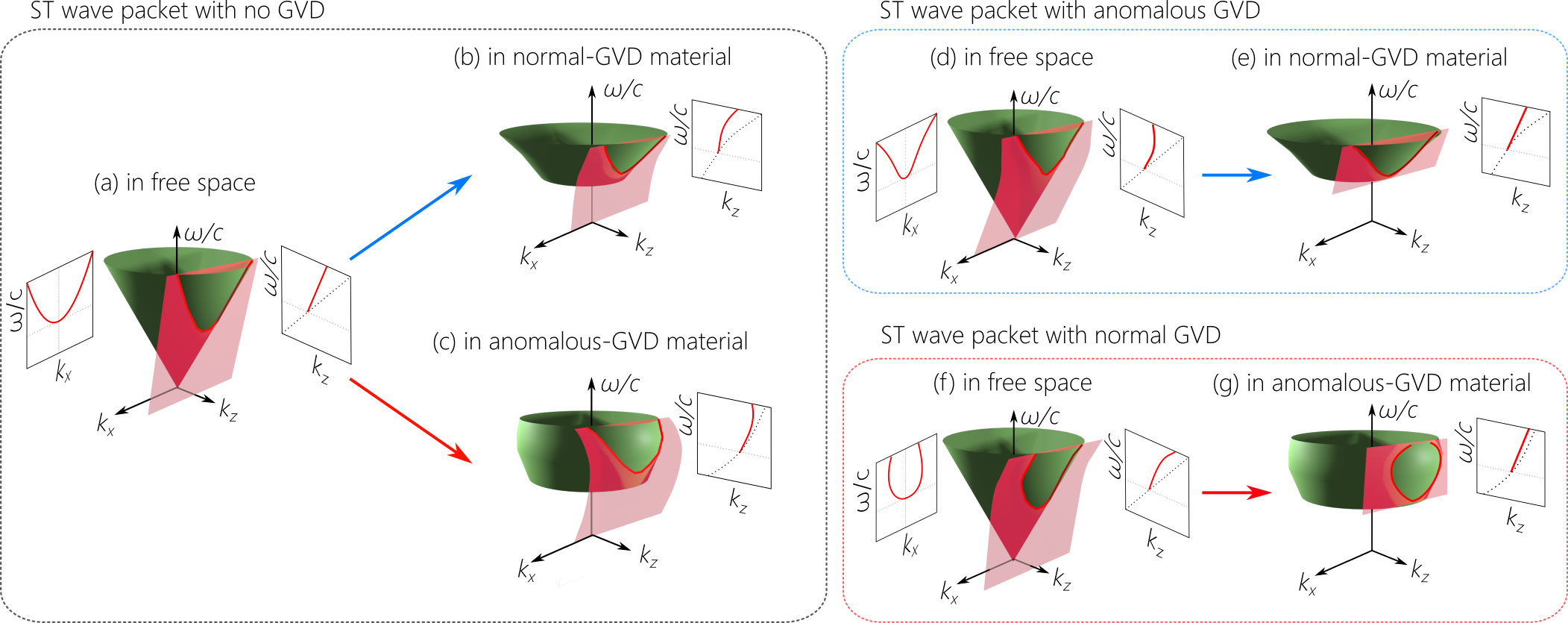}
\caption{(a) The spectral support domain of a propagation-invariant ST wave packet in free space on the surface of the light-cone $k_{x}^{2}+k_{z}^{2}\!=\!(\tfrac{\omega}{c})^{2}$, and its projections onto the $(k_{x},\tfrac{\omega}{c})$ and $(k_{z},\tfrac{\omega}{c})$ planes. The $(k_{z},\tfrac{\omega}{c})$-projection is a straight line, indicating absence of dispersion [Fig.~\ref{Fig:DiagramForNonDifferentiable}(a)]. (b) The ST wave packet from (a) after coupling to a normal-GVD or (c) an anomalous-GVD medium. The $(k_{z},\tfrac{\omega}{c})$-projections are now curved, indicating that the wave packet in the medium is dispersive. The $(k_{x},\tfrac{\omega}{c})$-projections are all the same in (a-c). (d) The spectral support domain for a ST wave packet endowed with \textit{anomalous} GVD in free space [Fig.~\ref{Fig:DiagramForNonDifferentiable}(b)]. The $(k_{z},\tfrac{\omega}{c})$-projection is no longer a straight line as in (a). (e) The ST wave packet from (d) after coupling to a medium with \textit{normal} GVD. The $(k_{z},\tfrac{\omega}{c})$-projection has been `straightened out', thus indicating dispersion-cancellation. (f) The spectral support domain for a ST wave packet endowed with \textit{normal} GVD in free space. (g) The ST wave packet from (f) after coupling to a medium with \textit{anomalous} GVD. Once again, The $(k_{z},\tfrac{\omega}{c})$-projection has been `straightened out'. In each case, the spectral projection onto the $(k_{x},\tfrac{\omega}{c})$-plane is invariant across the interface between free space and the dispersive medium.}
\label{Fig:Concept}
\end{figure*}

\subsection{Symmetrized anomalous and normal GVD in free space via angular dispersion}

The key to introducing anomalous \textit{or} normal GVD symmetrically into a pulsed field is to first change the group velocity along the propagation axis, which is given by: $\widetilde{v}_{\mathrm{a}}\!=\!c/\{\cos{\varphi_{\mathrm{o}}}-\omega_{\mathrm{o}}\varphi_{\mathrm{o}}^{(1)}\sin{\varphi_{\mathrm{o}}}\}$; where $\varphi_{\mathrm{o}}\!=\!\varphi_{\mathrm{a}}(\omega_{\mathrm{o}})$ and $\varphi_{\mathrm{o}}^{(1)}\!=\!\tfrac{d\varphi_{\mathrm{a}}}{d\omega}\big|_{\omega_{\mathrm{o}}}$ \cite{Porras03PRE2}. For on-axis propagation $\varphi_{\mathrm{o}}\!\rightarrow\!0$, $\widetilde{v}_{\mathrm{a}}\!\rightarrow\!c$ for any finite value of $\varphi_{\mathrm{o}}^{(1)}$. However, setting $\varphi_{\mathrm{a}}(\omega)\!=\!\varphi_{\mathrm{ST}}(\omega_{\mathrm{o}}+\Omega)\!\approx\!\eta\sqrt{\Omega/\omega_{\mathrm{o}}}$, which is \textit{not} differentiable at $\omega\!=\!\omega_{\mathrm{o}}$, yields a propagation-invariant ST wave packet with a group velocity $\widetilde{v}_{\mathrm{a}}\!=\!\tfrac{c}{\widetilde{n}_{\mathrm{a}}}$ and group index $\widetilde{n}_{\mathrm{a}}\!=\!1-\tfrac{1}{2}\eta^{2}$, where $\eta$ is a frequency-independent constant \cite{Hall21OL3NormalGVD,Hall22OEConsequences}. For a superluminal ST wave packet $\widetilde{v}_{\mathrm{a}}\!>\!c$, $\omega_{\mathrm{o}}$ is the minimum frequency in the spectrum [Fig.~\ref{Fig:DiagramForNonDifferentiable}(a)]; and for the subluminal counterpart $\widetilde{v}_{\mathrm{a}}\!<\!c$, $\omega_{\mathrm{o}}$ is the maximum [Fig.~\ref{Fig:DiagramForNonDifferentiable}(d)]. The axial wave numbers are now limited by the modified light-line $k_{z}(\omega)\!=\!\tfrac{\omega}{c}\cos{\varphi_{\mathrm{ST}}}\!=\!k_{\mathrm{o}}+\tfrac{\Omega}{\widetilde{v}_{\mathrm{a}}}$ rather than the free-space light-line $k_{z}\!=\!k_{\mathrm{o}}+\tfrac{\Omega}{c}$ [Fig.~\ref{Fig:DiagramForNonDifferentiable}(a,d)]. Because $k_{z}$ is linear in $\Omega$, the ST wave packet is dispersion-free.  

It can now be appreciated how anomalous \textit{or} normal GVD are induced in free space. Anomalous GVD requires reducing $k_{z}$ by further increasing the angular tilt, $\varphi_{\mathrm{anom}}(\omega)\!=\!\varphi_{\mathrm{ST}}(\omega)+\Delta\varphi(\omega)$, with respect to the GVD-free configuration [Fig.~\ref{Fig:DiagramForNonDifferentiable}(b,e)]. This regime is also accessible via conventional TPFs. In contrast, normal GVD requires increasing $k_{z}$ with respect to the GVD-free configuration by reducing the angular tilt $\varphi_{\mathrm{norm}}(\omega)\!=\!\varphi_{\mathrm{ST}}(\omega)-\Delta\varphi(\omega)$ [Fig.~\ref{Fig:DiagramForNonDifferentiable}(c,f)]. Because of the space that opened up above the free-space light-line $k_{z}\!=\!k_{\mathrm{o}}+\Omega/c$ but below the modified light-line $k_{z}\!=\!k_{\mathrm{o}}+\Omega/\widetilde{v}_{\mathrm{a}}$, we can increase $k_{z}$ without the field becoming evanescent. This regime is inaccessible to TPFs  [Fig.~\ref{Fig:DiagramFoDifferentiable}(h)] or other structured pulsed fields, and requires non-differentiable angular dispersion for its realization.

\subsection{Coupling from free space to a dispersive medium}

To analyze quantitatively the scenario illustrated in Fig.~\ref{Fig:DiagramForNonDifferentiable}, the representation of the spectral support domain of pulsed fields on the surface of the light-cone is a useful guide \cite{Donnelly93PRSLA,Yessenov22AOP}. In free space, the light-cone is $k_{x}^{2}+k_{z}^{2}\!=\!(\tfrac{\omega}{c})^{2}$, where $k_{x}$ is the transverse wave number or spatial frequency (we hold the field uniform along $y$ for simplicity). The spectral support domain for a propagation-invariant ST wave packet traveling at a group velocity $\widetilde{v}_{\mathrm{a}}$ is the intersection of the light-cone with a plane $k_{z}\!=\!k_{\mathrm{o}}+\Omega/\widetilde{v}_{\mathrm{a}}$ that is parallel to the $k_{x}$-axis and makes an angle $\theta_{\mathrm{a}}$ (the spectral tilt angle) with the $k_{z}$-axis \cite{Kondakci17NP,Kondakci19NC,Yessenov19PRA}, where $\widetilde{v}_{\mathrm{a}}\!=\!c\tan{\theta_{\mathrm{a}}}$ [Fig.~\ref{Fig:Concept}(a)]. The spectral projection onto the $(k_{z},\tfrac{\omega}{c})$-plane is a straight line, and onto the $(k_{x},\tfrac{\omega}{c})$-plane is a conic section that can be approximated by a parabola in the vicinity of $k_{x}\!=\!0$ in the paraxial regime \cite{Yessenov19OE,Yessenov19PRA}.

In presence of GVD due to chromatic dispersion, the dispersion relationship $k_{x}^{2}+k_{z}^{2}\!=\!(n\tfrac{\omega}{c})^{2}$ corresponds to a modified light-cone [Fig.~\ref{Fig:Concept}(b,c)]. At normal incidence on a planar interface, $k_{x}$ and $\Omega$ are invariant, so the $(k_{x},\tfrac{\omega}{c})$-projection is the same in free space and the dispersive medium:
\begin{equation}\label{Eq:GVDFreeInDispersiveMedium}
k_{x}^{2}=\underbrace{\frac{\omega^{2}}{c^{2}}\!-\!\left(k_{\mathrm{o}}\!+\!\frac{\Omega}{\widetilde{v}_{\mathrm{a}}}\right)^{2}}_{\mathrm{in\;free\;space}}\!=\!\underbrace{\left(n_{\mathrm{m}}k_{\mathrm{o}}\!+\!\frac{\Omega}{\widetilde{v}_{\mathrm{m}}}\!+\!\frac{1}{2}k_{2\mathrm{m}}\Omega^{2}\right)^{2}\!-k_{z}^{2}}_{\mathrm{in\;the\;medium}}. 
\end{equation}
However, the $(k_{z},\tfrac{\omega}{c})$-projection changes because the light-cone structure has been modified. The ST wave packet that was propagation-invariant in free space now experiences GVD in the dispersive medium. We expand the axial wave number in the medium as
$k_{z}\!\approx\!n_{\mathrm{m}}k_{\mathrm{o}}+\tfrac{\Omega}{\widetilde{v}}+\tfrac{1}{2}k_{2\mathrm{m}}'\Omega^{2}$, where $\widetilde{v}\!=\!\tfrac{c}{\widetilde{n}}$ is the group velocity of the ST wave packet in the medium (which need not be equal to $\widetilde{v}_{\mathrm{m}}$), and $k_{2\mathrm{m}}'$ is the effective GVD coefficient (which can differ from the GVD coefficient $k_{2\mathrm{m}}$ in the medium).

By equating the \textit{first}-order $\Omega$ terms in Eq.~\ref{Eq:GVDFreeInDispersiveMedium}, we obtain:
\begin{equation}\label{Eq:RefractionLaw}
1-\widetilde{n}_{\mathrm{a}}=n_{\mathrm{m}}(\widetilde{n}_{\mathrm{m}}-\widetilde{n}),
\end{equation}
which can be recognized as the law of refraction for a ST wave packet in a dispersive medium derived in \cite{He21Arxiv,Yessenov22OLDispersiveRefraction} that governs the change in group velocity from $\widetilde{v}_{\mathrm{a}}\!=\!\tfrac{c}{\widetilde{n}_{\mathrm{a}}}$ in free space to $\widetilde{v}\!=\!\tfrac{c}{\widetilde{n}}$ in the medium. Indeed, the quantity $n_{\mathrm{m}}(\widetilde{n}_{\mathrm{m}}-\widetilde{n})$ is a refractive invariant for ST wave packets at normal incidence on planar interfaces between dispersive media, which we have called the `spectral curvature' because it is related to the curvature of the parabolic $(k_{x},\tfrac{\omega}{c})$-projection in the vicinity of $k_{x}\!=\!0$. This relationship indicates that a subluminal ST wave packet in free space $\widetilde{v}_{\mathrm{a}}\!<\!c$ remains subluminal in the medium $\widetilde{v}\!<\!\widetilde{v}_{\mathrm{m}}$, and similarly for superluminal wave packets.

Equating the \textit{second}-order $\Omega^{2}$ terms in Eq.~\ref{Eq:GVDFreeInDispersiveMedium} yields a relationship between the GVD coefficient of the medium $k_{2\mathrm{m}}$ and the effective GVD coefficient experienced by the wave packet $k_{2\mathrm{m}}'$:
\begin{equation}
k_{2\mathrm{m}}'=k_{2\mathrm{m}}+\frac{1}{n_{\mathrm{m}}}\frac{\Delta}{c\omega_{\mathrm{o}}}.
\end{equation}
From this we conclude that the ST wave packet experiences the normal or anomalous GVD intrinsic to the medium itself [Fig.~\ref{Fig:Concept}(b,c)] except for an offset term:
\begin{equation}\label{Eq:offsetDelta}
\Delta=(\widetilde{n}_{\mathrm{m}}^{2}-\widetilde{n}^{2})-(1-\widetilde{n}_{\mathrm{a}}^{2}).
\end{equation}
In most cases where the deviation from the luminal limit is small ($\widetilde{n}_{\mathrm{a}}\!\rightarrow\!1$ and $\widetilde{n}\!\rightarrow\!\widetilde{n}_{\mathrm{m}}$), $\Delta$ can be ignored, and we have $k_{2\mathrm{m}}'\!\approx\!k_{2\mathrm{m}}$.

\subsection{Achieving dispersion-free propagation in presence of GVD}

Dispersion-free propagation in the dispersive medium is achieved by modifying the structure of the ST wave packet in free space to introduce GVD of opposite sign to that of the medium: cancelling normal GVD necessitates endowing the ST wave packet in free space with anomalous GVD [Fig.~\ref{Fig:Concept}(d,e)], and vice versa [Fig.~\ref{Fig:Concept}(f,g)]. In free space, the $(k_{z},\tfrac{\omega}{c})$-projection is no longer a straight line, but rather takes the form $k_{z}\!=\!k_{\mathrm{o}}+\tfrac{\Omega}{\widetilde{v}_{\mathrm{a}}}+\tfrac{1}{2}k_{2\mathrm{a}}\Omega^{2}$, where $k_{2\mathrm{a}}$ is the GVD coefficient introduced into the ST wave packet. The spectral support domain on the free-space light-cone is its intersection with a planar \textit{curved} surface that is parallel to the $k_{x}$-axis:
\begin{equation}\label{Eq:FreeSpaceDispersionEquation}
k_{x}^{2}\!+\!\left(k_{\mathrm{o}}+\frac{\Omega}{\widetilde{v}_{\mathrm{a}}}+\frac{1}{2}k_{2\mathrm{a}}\Omega^{2}\right)^{2}=\left(k_{\mathrm{o}}+\frac{\Omega}{c}\right)^{2}.
\end{equation}
The change in the light-cone structure in the medium in conjunction with the invariance of the $(k_{x},\tfrac{\omega}{c})$-projection can yield a $(k_{z},\tfrac{\omega}{c})$-projection in the medium that is a \textit{straight line}. In other words, the curved $(k_{z},\tfrac{\omega}{c})$-projection in free space has been `straightened out' in the medium such that $k_{z}\!=\!n_{\mathrm{m}}k_{\mathrm{o}}+\tfrac{\Omega}{\widetilde{v}}$, thus signifying dispersion-free propagation in the medium:
\begin{equation}\label{Eq:DispersiveMediumEquation}
k_{x}^{2}\!+\!\left(n_{\mathrm{m}}k_{\mathrm{o}}+\frac{\Omega}{\widetilde{v}}\right)^{2}=\left(n_{\mathrm{m}}k_{\mathrm{o}}+\frac{\Omega}{\widetilde{v}_{\mathrm{m}}}+\frac{1}{2}k_{2\mathrm{m}}\Omega^{2}\right)^{2}.
\end{equation}

Equating the $\Omega$-terms in Eq.~\ref{Eq:FreeSpaceDispersionEquation} and Eq.~\ref{Eq:DispersiveMediumEquation} yields $1-\widetilde{n}_{\mathrm{a}}\!=\!n_{\mathrm{m}}(\widetilde{n}_{\mathrm{m}}-\widetilde{n})$ as in Eq.~\ref{Eq:RefractionLaw}), whereas equating the $\Omega^{2}$-terms yields:
\begin{equation}\label{Eq:DispersionCancellationGVD}
k_{2\mathrm{a}}=-n_{\mathrm{m}}k_{2\mathrm{m}}-\frac{\Delta}{c\omega_{\mathrm{o}}}.
\end{equation}
That is, dispersion cancellation requires that the dispersion coefficient introduced in free space $k_{2\mathrm{a}}$ must have the opposite sign to that of the medium $k_{2\mathrm{m}}$, and to be weighted by the refractive index $n_{\mathrm{m}}$. This result is similar to that for a TPF (Appendix) except that the GVD to be cancelled can be \textit{either} normal \textit{or} anomalous (in addition to the minor offset term $\Delta$).

\section{Experiment}

\subsection{Spatio-temporal spectral synthesis}

\begin{figure}[t!]
\centering
\includegraphics[width=8.6cm]{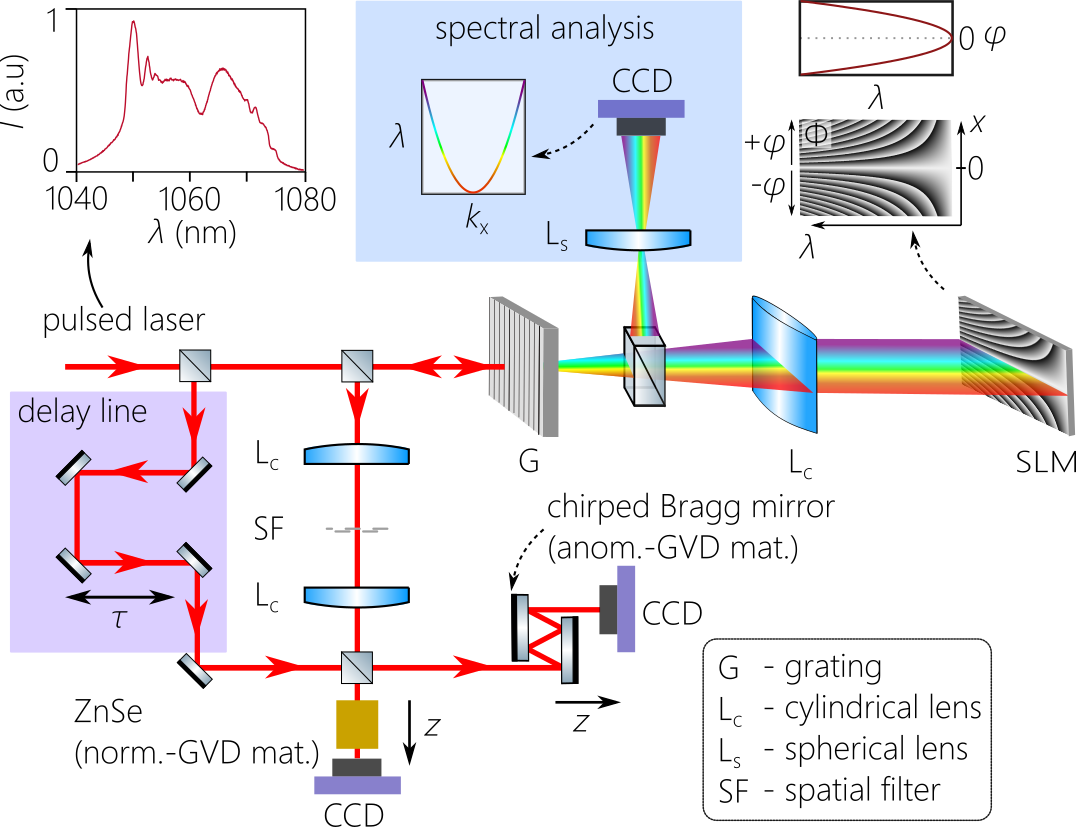}
\caption{Schematic of the setup for synthesizing and characterizing dispersive ST wave packets. The inset in the upper left corner is the spectrum of the pulsed laser used. The insets in the upper right corner depict the two-dimensional phase pattern $\Phi$ imparted by the SLM to the spectrally resolved wave front, and the resulting angular dispersion $\varphi(\lambda)$ inculcated into the incident field.}
\label{Fig:Setup}
\end{figure}

To synthesize dispersive ST wave packets in free space, we make use of the universal angular-dispersion synthesizer described in \cite{Hall21OEUniversal} and depicted in Fig.~\ref{Fig:Setup}. We start with plane-wave femtosecond pulses of width $\approx\!100$~fs and bandwidth $\approx\!25$~nm at a central wavelength of $\lambda_{\mathrm{o}}\!\approx\!1064$~nm (Spark Lasers; Alcor). The pulses are spectrally resolved via a diffraction grating (1200~lines/mm) followed by a collimating cylindrical lens of focal length $f\!=\!500$~mm. At the focal plane of the lens we place a reflective, phase-only spatial light modulator (SLM; Meadowlark, E19X12) that imparts a 2D phase distribution to the impinging spectrally resolved wave front. Each wavelength occupies a column on the SLM, along which we impose the phase $\Phi(x,\lambda)\!=\!\pm\tfrac{2\pi}{\lambda}\sin{\{\phi(\lambda)\}}x$, where $\varphi(\lambda)$ is the deflection angle for $\lambda$ with respect to the $z$-axis. An example of the phase pattern $\Phi(x,,\lambda)$ is depicted in Fig.~\ref{Fig:Setup}, inset. The upper half of each SLM column deflects the wavelength $\lambda$ at an angle $\varphi(\lambda)$, while the lower half deflects it at $-\varphi(\lambda)$. This yields a symmetric spatio-temporal angular spectrum $\varphi(\lambda)$ [Fig.~\ref{Fig:Setup}, inset], and produces a X-shaped wave-packet profile, whereas that for TPFs [Fig.~\ref{Fig:DiagramFoDifferentiable}(f)] comprises one branch of the X-shaped profile. The wave front retro-reflected from the SLM returns to the grating where the ST wave packet is formed, and the dispersive sample is placed in its path.

\begin{figure*}[t!]
\centering
\includegraphics[width=13cm]{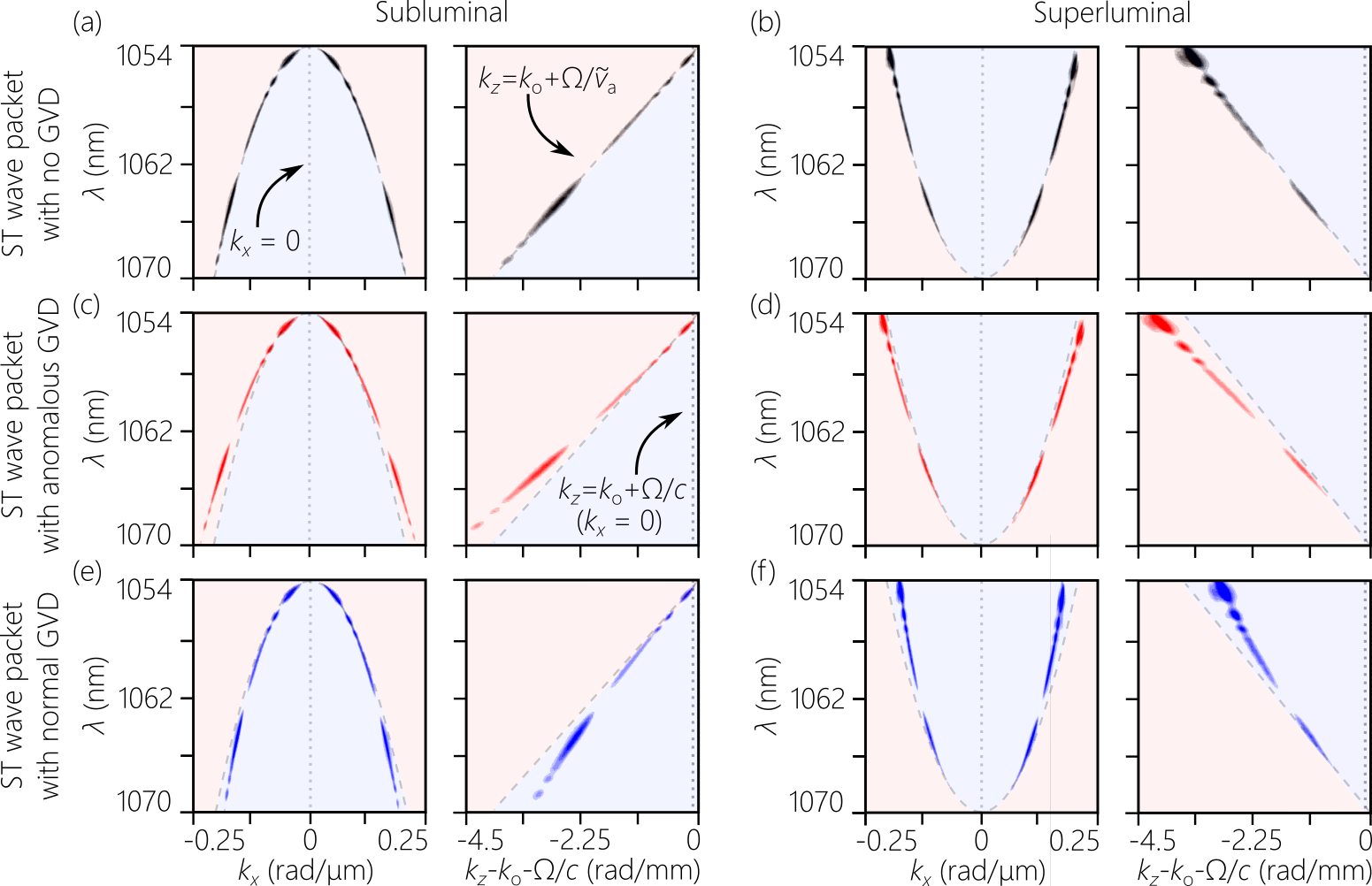}
\caption{Measured spectral projections onto the $(k_{x},\lambda)$ and $(k_{z},\lambda)$ planes, (a,c,e) for subluminal wave packets with $\theta_{\mathrm{a}}\!=\!44^{\circ}$ and $\widetilde{v}_{\mathrm{a}}\!=\!0.96c$, and (b,d,f) for superluminal wave packets with $\theta_{\mathrm{a}}\!=\!46^{\circ}$ and $\widetilde{v}_{\mathrm{a}}\!=\!1.04c$. (a,b) Dispersion-free ST wave packets in free space; (c,d) ST wave packets endowed with anomalous GVD in free space, $c\omega_{\mathrm{o}}k_{2\mathrm{a}}\!=\!-0.8$; and (e,f) ST wave packets endowed with anomalous GVD in free space, $c\omega_{\mathrm{o}}k_{2\mathrm{a}}\!=\!0.8$. The dotted curves in the $(k_{x},\lambda)$-plane are the theoretical spectra of the GVD-free wave-packets in (a) and (b), and are a guide for the eye in the case of their dispersive counterparts. The dashed lines in the $(k_{z},\lambda)$-plane are the modified light-lines $k_{z}\!=\!k_{\mathrm{o}}+\tfrac{\Omega}{\widetilde{v}_{\mathrm{a}}}$. The free-space light-line is the vertical axis at $k_{z}-k_{\mathrm{o}}-\tfrac{\Omega}{c}\!=\!0$.}
\label{Fig:MeasuredSpectra}
\end{figure*}

\subsection{Dispersive samples}

The normal-GVD medium is ZnSe (Thorlabs; WG71050) formed of multiple 1-inch-diameter discs of thickness 5~mm each, stacked to a maximum thickness of 30~mm. Using the Sellmeier equation for ZnSe, $n^{2}(\lambda)\!=\!4+\tfrac{1.9\lambda^{2}}{\lambda^{2}-0.113}$ ($\lambda$ in units of $\mu$m) \cite{Marple64JAP}, we obtain at $\lambda_{\mathrm{o}}\!=\!1064$~nm an index $n_{\mathrm{m}}\!\approx\!2.49$, group index $\widetilde{n}_{\mathrm{m}}\!\approx\!2.57$, and GVD parameter $k_{2\mathrm{m}}\!\approx\!+607.45$~fs$^2$/mm. It is useful to exploit the dimensionless GVD parameter $c\omega_{\mathrm{o}}k_{2\mathrm{m}}\!\approx\!0.32$.

The anomalous-GVD sample comprises a pair of chirped Bragg mirrors (Edmund Optics, 12-335) that generate -1000 fs$^2$ group delay dispersion (GDD) per reflection. By changing the distance separating the two mirrors, we can increase the number of reflections for a given propagation distance before the wave packet emerges, thus controllably increasing the GDD from -2000 to -15000 fs$^2$. The GVD is then taken to be the GDD divided by the total length propagated at a fixed incident angle of $7^{\circ}$, resulting in an effective GVD coefficient of $k_{2\mathrm{m}}\!\approx\!-500$~fs$^{2}$/mm and a medium length extending up to 30~mm. The dimensionless GVD parameter here is $c\omega_{\mathrm{o}}k_{2\mathrm{m}}\!\approx\!-0.25$, and $n_{\mathrm{m}}\!\approx\!\widetilde{n}_{\mathrm{m}}\!\approx\!1$ because the ST wave packet travels predominantly in free space. 

\section{Spectral measurements}

The spatio-temporal spectrum projected onto the $(k_{x},\lambda)$-plane is obtained after implementing a spatial Fourier transform on the spectrally resolved wave front reflecting back from the SLM [Fig.~\ref{Fig:Setup}]. The intensity distribution is then recorded by a CCD camera. The result is a parabola centered at $k_{x}\!=\!0$ of spatial bandwidth $\Delta k_{x}\!\approx\!0.23$~rad/$\mu$m (corresponding to a spatial width of $\Delta x\!\approx\!20$~$\mu$m at the pulse center) and a temporal bandwidth of $\Delta\lambda\!\approx\!16$~nm (corresponding to a temporal linewidth of $\Delta T\!\approx\!200$~fs at the beam center). We plot in Fig.~\ref{Fig:MeasuredSpectra} the measured spectra for two classes of ST wave packets: subluminal in Fig.~\ref{Fig:MeasuredSpectra}(a,c,e) with $\theta_{\mathrm{a}}\!=\!44^{\circ}$ and $\widetilde{v}_{\mathrm{a}}\!\approx\!0.96c$, and superluminal in Fig.~\ref{Fig:MeasuredSpectra}(b,d,f) with $\theta_{\mathrm{a}}\!=\!46^{\circ}$ and $\widetilde{v}_{\mathrm{a}}\!\approx\!1.04c$. In each category we produce three distinct wave packets in free space: (1) a GVD-free wave packet that is propagation invariant [Fig.~\ref{Fig:MeasuredSpectra}(a,b)]; (2) a dispersive ST wave packet endowed with \textit{anomalous} GVD $k_{2\mathrm{a}}\!<\!0$ [Fig.~\ref{Fig:MeasuredSpectra}(c,d)]; and (3) a dispersive ST wave packet endowed with \textit{normal} GVD $k_{2\mathrm{a}}\!>\!0$ in free space [Fig.~\ref{Fig:MeasuredSpectra}(e,f)]. In all cases, each spatial frequency $k_{x}$ is associated with a single wavelength $\lambda$. After introducing anomalous GVD $k_{2\mathrm{a}}\!<\!0$, $|k_{x}|$ \textit{increases} with respect to its GVD-free counterpart, corresponding to the required increase in propagation angle [Fig.~\ref{Fig:DiagramForNonDifferentiable}(b,e)]. Alternatively, $|k_{x}|$ \textit{decreases} with respect to the GVD-free wave packet after incorporating normal GVD $k_{2\mathrm{a}}\!>\!0$ [Fig.~\ref{Fig:MeasuredSpectra}(e,f)] corresponding to the required decrease in propagation angle [Fig.~\ref{Fig:DiagramForNonDifferentiable}(c,f)].

From these spectra in the $(k_{x},\lambda)$-plane, we extract the spectral projection onto the $(k_{z},\lambda)$-plane for the field in free space, making use of the relationship $k_{z}^{2}\!=\!(\tfrac{\omega}{c})^{2}-k_{x}^{2}$. This spectral projection for the GVD-free wave packets is a straight line [Fig.~\ref{Fig:MeasuredSpectra}(a,b)]. These wave packets propagate invariantly in free space. However, the $(k_{z},\lambda)$-projections curve away from that straight line in presence of GVD [Fig.~\ref{Fig:MeasuredSpectra}(c-f)]. The dashed lines in the $(k_{z},\lambda)$-plane [Fig.~\ref{Fig:MeasuredSpectra}(c-f)] are the modified light-lines $k_{z}\!=\!k_{\mathrm{o}}+\Omega/\widetilde{v}_{\mathrm{a}}$. Introducing anomalous GVD reduces $k_{z}$, whereas incorporating normal GVD increases $k_{z}$ at each wavelength, which is consistent with our goal as illustrated in Fig.~\ref{Fig:DiagramForNonDifferentiable}. Therefore, the measurements confirm that the targeted spatio-temporal spectra have indeed been produced in free space. We now proceed to verify that the expected propagation dynamics is produced in free space and in the dispersive media.

\section{Dispersive space-time wave packets in free space}

We reconstruct the spatio-temporal envelope of the wave-packet intensity profile $I(x,z;\tau)$ at a given axial plane $z$ via linear interferometry making use of the initial laser pulses as a reference \cite{Kondakci19NC}; see Fig.~\ref{Fig:Setup}. When the ST wave packet and the reference pulse overlap in space and time, spatially resolved fringes are recorded by a CCD camera whose visibility is used to reconstruct the wave packet profile as an optical delay $\tau$ is swept in the path of the reference pulse. Furthermore, the profiles of the ST wave packet at different axial planes $z$ are reconstructed by displacing the CCD camera to the target plane $z$ and compensating for the relative group delay between the ST wave packet (travelling at $\widetilde{v}_{\mathrm{a}}\!=\!c\tan{\theta_{\mathrm{a}}}$) and the reference pulse (travelling at $c$).

We start off with a subluminal ($\theta_{\mathrm{a}}\!=\!44^{\circ}$) propagation-invariant ST wave packet in free space, and plot in Fig.~\ref{Fig:DataProfiles}(a) the measured profile $I(x,z;\tau)$ at three axial planes in free space ($z\!=\!0$, 15, and 30~mm) reconstructed in a frame traveling at $\widetilde{v}_{\mathrm{a}}$. The X-shaped ST wave packet travels invariantly without distortion. The on-axis pulsewidth is constant in free space at $\Delta T\!\approx\!200$~fs. However, once this wave packet is coupled to a dispersive medium, pulse broadening is observed; see Fig.~\ref{Fig:DataProfiles}(b) for the normal-GVD medium and Fig.~\ref{Fig:DataProfiles}(c) for its anomalous-GVD counterpart. The pulsewidth increases monotonically from $\Delta T\!\approx\!200$~fs to $\Delta T\!\approx\!600$~fs after 30~mm in either medium [Fig.~\ref{Fig:DataProfiles}(d,e)].

Crucially, accompanying pulse broadening is an asymmetry between the structure of the wave packets in the normal- and anomalous-GVD media in regards to the direction of pulse broadening. In a normal-GVD medium, the pulse broadens towards later delays with respect to $\tau\!=\!0$, whereas it boadens towards advanced delays. The distinct field structures that emerge in these two cases allow us to unambiguously delineate the wave packet at the output of the anomalous- and normal-GVD media. In Fig.~\ref{Fig:DataProfiles}(d) we plot the on-axis pulse profiles $I(0,z;\tau)$ at $z\!=\!0$, 15~mm, and 30~mm for the wave packets in Fig.~\ref{Fig:DataProfiles}(a-c) to highlight this asymmetry, which provides a clear signature of the type of GVD experienced by the wave packet. Note, however, that the rate of increase in pulsewidth $\Delta T$ with distance [Fig.~\ref{Fig:DataProfiles}(e)] does \textit{not} depend on the sign of the GVD \cite{SalehBook07}.

\section{Propagation invariance in dispersive media}

\subsection{Normal-GVD cancellation}

\begin{figure}[t!]
\centering
\includegraphics[width=8.6cm]{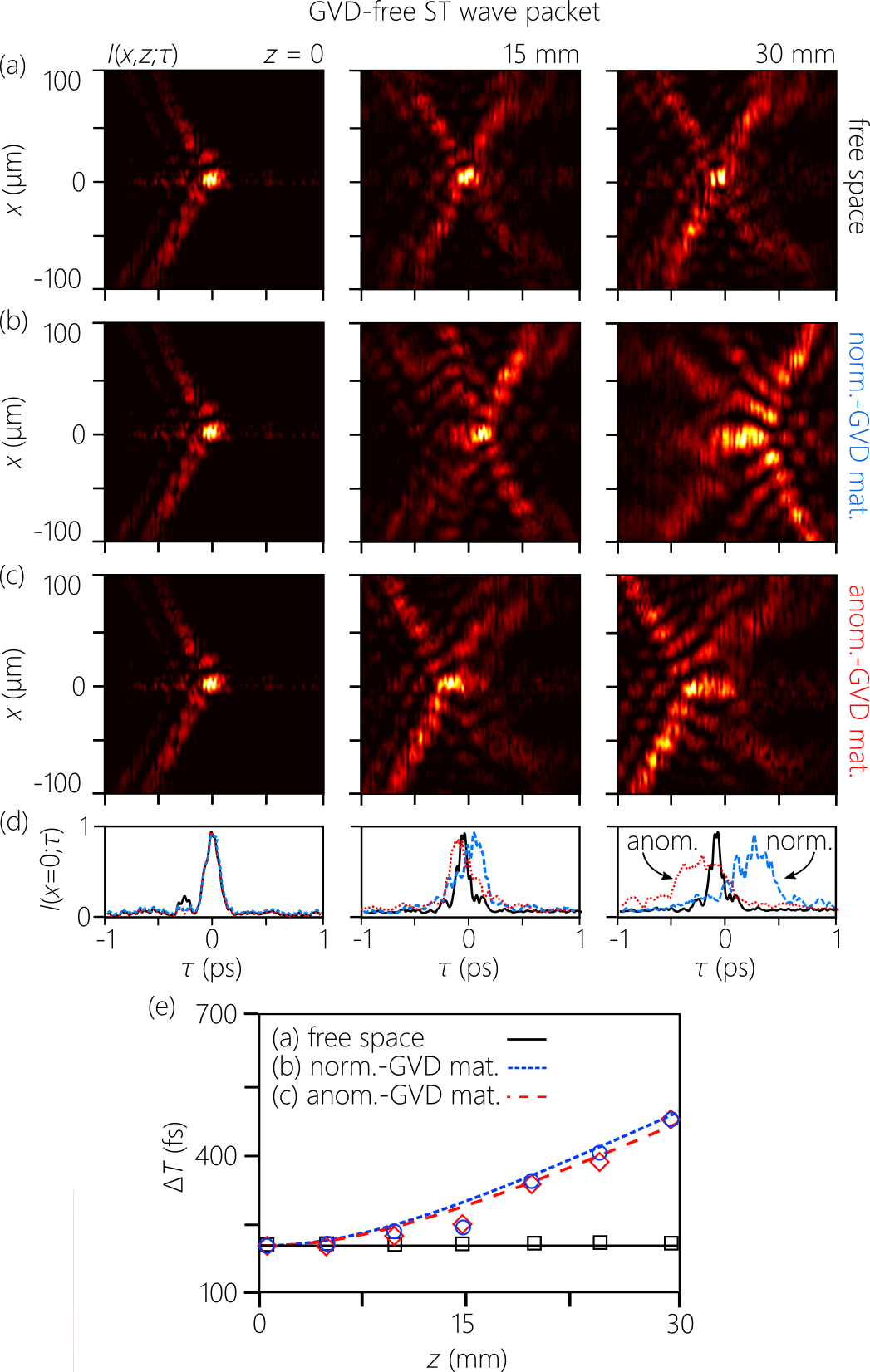}
\caption{(a-c) Measured spatio-temporal intensity profiles $I(x,z;\tau)$ at $z\!=\!0$, 15~mm, and 30~mm for a ST wave packet (a) in free space, (b) in a normal-GVD medium, and (c) in an anomalous-GVD medium. The ST wave packet is propagation invariant in free space (a), and thus experiences GVD in the dispersive media (b,c). (d) On-axis $x\!=\!0$ profiles $I(0,z;\tau)$ for the ST wave packets in (a-c). The three panels provide the pulse profiles at $z\!=\!0$ (where all three coincide), $z\!=\!15$~mm, and $z\!=\!30$~mm (where the pulses in the dispersive media have dispersed). (e) On-axis pulsewidth $\Delta T$ measured at 5-mm axial intervals}
\label{Fig:DataProfiles}
\end{figure}

\begin{figure*}[t!]
\centering
\includegraphics[width=17.6cm]{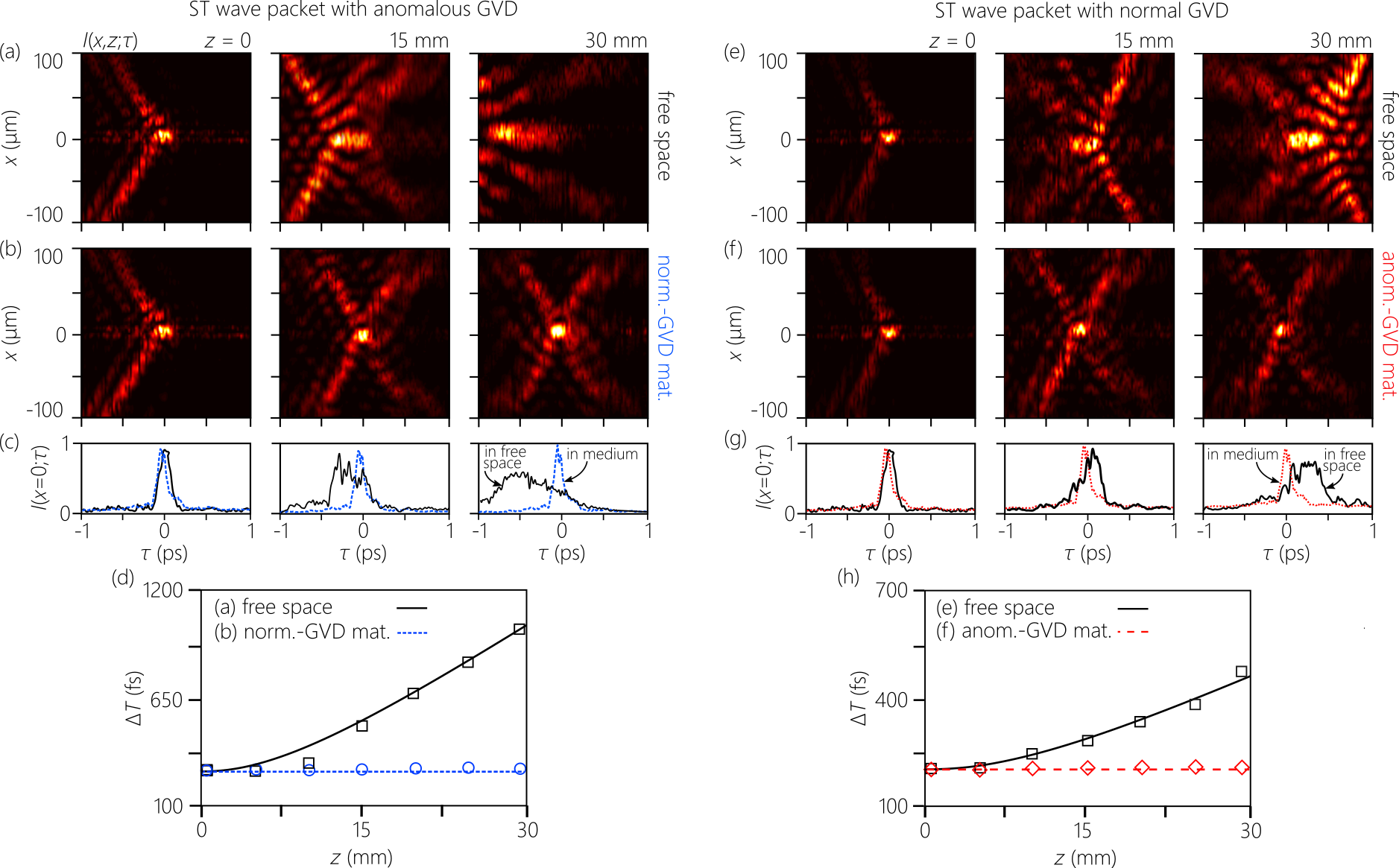}
\caption{Cancellation of either normal or anomalous GVD. (a) Measured spatio-temporal intensity profiles $I(x,z;\tau)$ at $z\!=\!0$, 15~mm, and 30~mm for a ST wave packet endowed with anomalous GVD in free space, and (b) in a normal-GVD medium, whereupon dispersion is cancelled. (c) On-axis $x\!=\!0$ profiles $I(0,z;\tau)$ for the ST wave packets in (a,b). The panels provide the pulse profiles at $z\!=\!0$ (where the two coincide), $z\!=\!15$~mm, and $z\!=\!30$~mm (where the pulse has dispersed in free space but not in the medium). (d) On-axis pulsewidth $\Delta T$ measured at 5-mm axial intervals. (e) Measured spatio-temporal intensity profiles $I(x,z;\tau)$ at $z\!=\!0$, 15~mm, and 30~mm for a ST wave packet endowed with normal GVD in free space, and (f) in an anomalous-GVD medium, whereupon dispersion is cancelled. (g,h) Same as (c,d) but for the ST wave packets in (e,f).}
\label{Fig:DataProfilesNormalMaterialGVD}
\end{figure*}

GVD-free propagation in ZnSe in the normal-GVD regime requires introducing anomalous GVD into the ST wave packet in free space. We set $k_{2\mathrm{a}}\!\approx\!-1500$~fs$^{2}$/mm ($c\omega_{\mathrm{o}}k_{2\mathrm{m}}\!\approx\!-0.8$) for a subluminal ST wave packet and monitor its propagation in free space, whereupon it exhibits dispersive temporal broadening [Fig.~\ref{Fig:DataProfilesNormalMaterialGVD}(a)]. Moreover, the temporal asymmetry exhibited by the wave packet confirms that it experiences anomalous GVD in free space; compare Fig.~\ref{Fig:DataProfilesNormalMaterialGVD}(a) to Fig.~\ref{Fig:DataProfiles}(c). However, once the wave packet is coupled to ZnSe, this behavior is halted, and the wave packet travels GVD-free with a propagation-invariant spatio-temporal profile independently of the distance up to a 30-mm-thick ZnSe sample [Fig.~\ref{Fig:DataProfilesNormalMaterialGVD}(b)]. The on-axis pulse broadening in free space depicted in Fig.~\ref{Fig:DataProfilesNormalMaterialGVD}(c,d) is in quantitative agreement with the expectation based on the GVD coefficient introduced, whereas the pulsewidth is constant after GVD-cancellation.

\subsection{Anomalous-GVD cancellation}

Propagation invariance in the anomalous-GVD sample requires introducing normal GVD into the ST wave packet in free space. Setting $k_{2\mathrm{a}}\!\approx\!500$~fs$^{2}$/mm ($c\omega_{\mathrm{o}}k_{2\mathrm{m}}\!\approx\!0.25$) and monitoring the propagation of the dispersive ST wave packet in free space reveals dispersive temporal broadening [Fig.~\ref{Fig:DataProfilesNormalMaterialGVD}(e)]. The temporal asymmetry in the wave packet spreading is consistent with normal GVD [compare Fig.~\ref{Fig:DataProfilesNormalMaterialGVD}(e) to Fig.~\ref{Fig:DataProfiles}(b)]. However, after traversing the chirped mirrors, the wave packet travels GVD-free with a propagation-invariant spatio-temporal profile independently of the sample thickness [Fig.~\ref{Fig:DataProfilesNormalMaterialGVD}(f)]. Once again, the broadening in the on-axis pulsewidth $\Delta T$ is in quantitative agreement with the expectation based on the GVD coefficient introduced, whereas the pulsewidth in the medium is constant after GVD-cancellation [Fig.~\ref{Fig:DataProfilesNormalMaterialGVD}(g,h)].

\begin{figure}[t!]
\centering
\includegraphics[width=8.6cm]{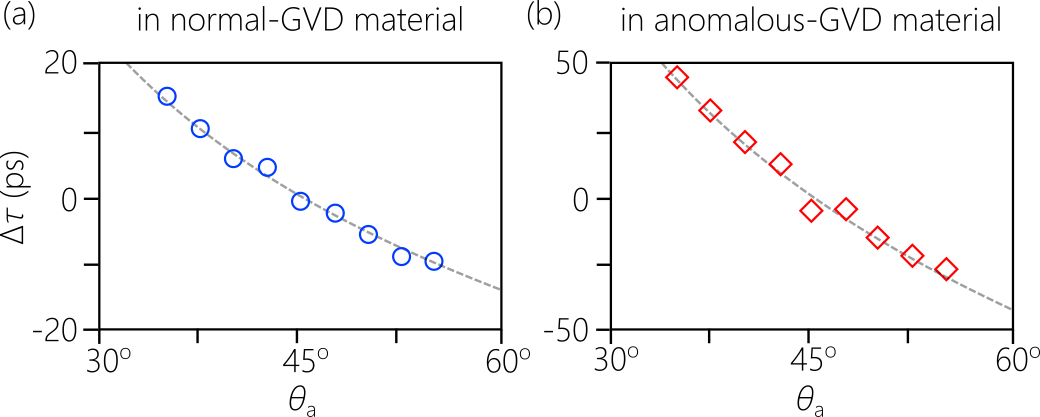}
\caption{(a) Measured group delay $\Delta\tau$ for a ST wave packet with respect to a conventional pulsed plane-wave upon traversing a fixed-thickness medium ($L\!=\!30$~mm) while tuning the free-space spectral tilt angle $\theta_{\mathrm{a}}$. This wave packet is endowed with anomalous GVD in free space and is coupled to a normal-GVD medium where it is GVD-free. (b) Same as (a) but for a ST wave packet endowed with normal GVD in free space traversing a fixed-thickness medium having anomalous GVD. The dashed curves in (a,b) are the theoretical predictions $\Delta\tau\!=\!\tfrac{L}{c}\widetilde{n}$, where $\widetilde{n}$ is determined from Eq.~\ref{Eq:RefractionLaw}.}
\label{Fig:DataDelays}
\end{figure}

\subsection{Independence of the group velocity and GVD-cancellation}

The measurements in Fig.~\ref{Fig:DataProfilesNormalMaterialGVD} were carried out at a fixed group velocity $\widetilde{v}_{\mathrm{a}}$ in free space. In many nonlinear optical applications that benefit from GVD-cancellation, it is also useful to also control the wave-packet group velocity. This can help group-velocity matching between pulses at disparate wavelengths while exploiting long crystals. In our scheme, the angular dispersion profile $\varphi(\lambda)$ can be controlled almost arbitrarily [Fig.~\ref{Fig:Setup}]: each wavelength $\lambda$ can be assigned a propagation angle $\varphi(\lambda)$ independently of all other wavelengths, and we can thus tune $\widetilde{v}_{\mathrm{a}}$ and $k_{2\mathrm{a}}$ independently \cite{Yessenov21ACSP}. We demonstrate this capability in Fig.~\ref{Fig:DataDelays} where we measure the group delay in normal- and anomalous-GVD samples of \textit{fixed} length ($L\!=\!30$~mm) while varying the spectral tilt angle $\theta_{\mathrm{a}}$ in free space. This results in tuning of the free-space group velocity $\widetilde{v}_{\mathrm{a}}\!=\!c\tan{\theta_{\mathrm{a}}}$ and hence the group velocity $\widetilde{v}$ in the medium (Eq.~\ref{Eq:RefractionLaw}). Throughout, we maintain GVD-cancellation; that is, the wave packet is invariant after the sample \textit{independently} of $\widetilde{v}_{\mathrm{a}}$. As $\widetilde{v}_{\mathrm{a}}$ is increased continuously from the subluminal regime ($\theta_{\mathrm{a}}\!<\!45^{\circ}$, $\widetilde{v}_{\mathrm{a}}\!<\!c$) to the superluminal ($\theta_{\mathrm{a}}\!>\!45^{\circ}$, $\widetilde{v}_{\mathrm{a}}\!>\!c$) regime, the group velocity of the wave packet in the medium $\widetilde{v}$ also increases, resulting in a concomitant drop in the group delay over the fixed sample length. This confirms that GVD-cancellation and group-velocity-tunability can be maintained independently of each other. 

\section{Inverting the group-velocity dispersion}

\begin{figure}[t!]
\centering
\includegraphics[width=8.6cm]{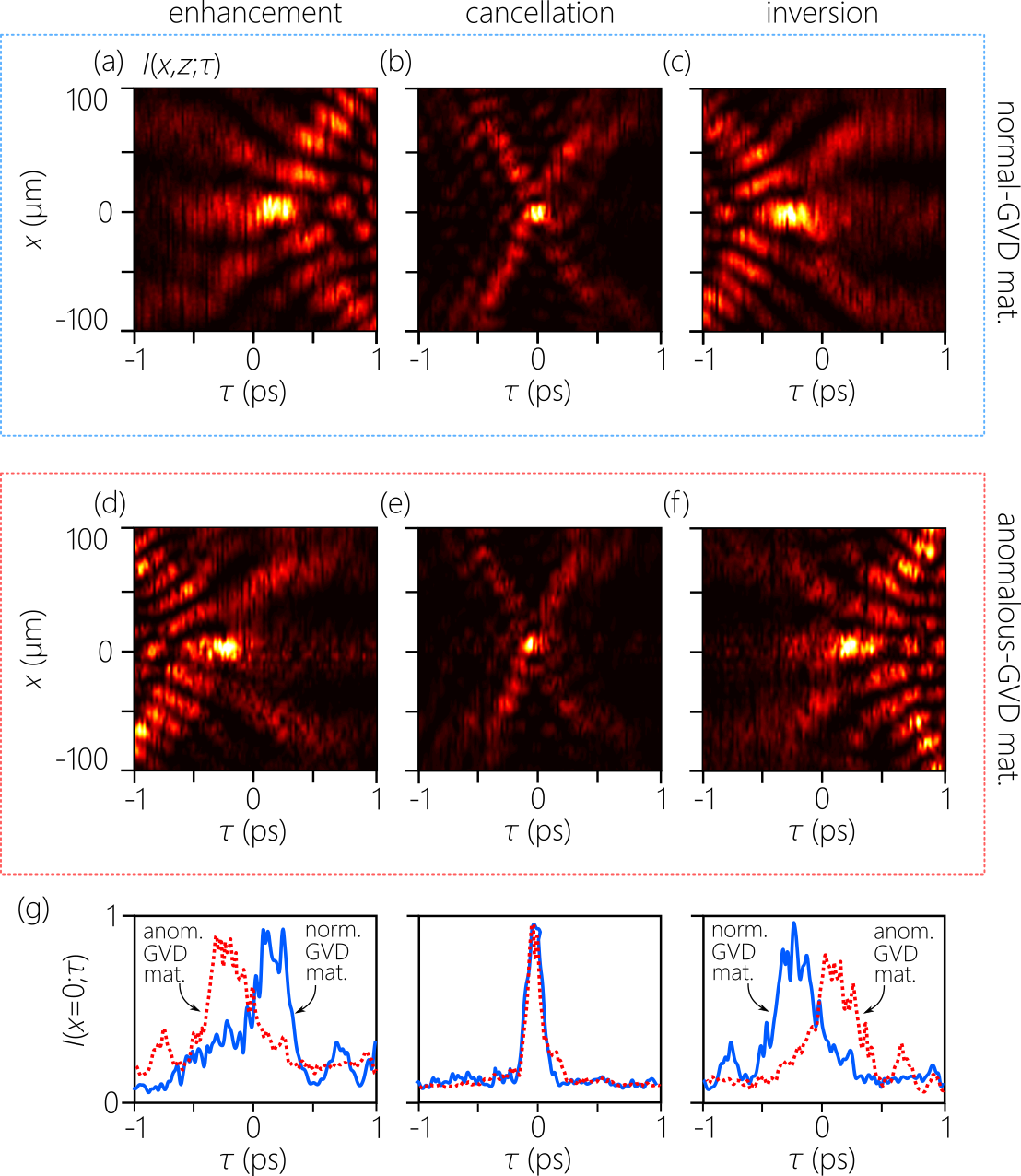}
\caption{Enhancing, eliminating, and inverting GVD in a dispersive medium by varying the GVD introduced into the ST wave packet in free space . (a-c) Spatio-temporal intensity profiles at $z\!=\!30$~mm in ZnSe (normal GVD) while varying the free-space GVD: (a) $c\omega_{\mathrm{o}}k_{2\mathrm{a}}\!=\!0.8$, (b) $-0.8$, and (c) $-2.25$. (d-f) Same as (a-c) but for the anomalous-GVD sample: (d) $c\omega_{\mathrm{o}}k_{2\mathrm{a}}\!\approx\!-0.25$, (e) $0.25$, and (f) $0.75$. In (a,d) the GVD experienced by the ST wave packets is enhanced; in (b,e) the GVD is cancelled; and in (c,f) the GVD is inverted.}
\label{Fig:DataReversal}
\end{figure}

When GVD-cancellation is \textit{not} achieved, the GVD coefficient in the medium $k_{2\mathrm{m}}'$ is given by:
\begin{equation}
k_{2\mathrm{m}}'=k_{2\mathrm{m}}+\frac{k_{2\mathrm{a}}}{n_{\mathrm{m}}}+\frac{1}{n_{\mathrm{m}}}\frac{\Delta}{c\omega_{\mathrm{o}}}.
\end{equation}
In other words, the effective GVD in the medium $k_{2\mathrm{m}}'$ combines the intrinsic material GVD $k_{2\mathrm{m}}$ (due to chromatic dispersion) with the free-space GVD introduced into the ST wave packet (via non-differentiable angular dispersion), in addition to the negligible offset $\Delta$. By varying $k_{2\mathrm{a}}$ we can realize one of three different scenarios. First, the GVD introduced in free space can reinforce the GVD in the medium ($k_{2\mathrm{a}}$ has the same sign as $k_{2\mathrm{m}}$), leading to GVD \textit{enhancement} ($|k_{2\mathrm{m}}'|\!>\!|k_{2\mathrm{m}}|$). We present measurements for such a scenario in Fig.~\ref{Fig:DataReversal}(a,d) where we enhance the dispersion in the normal-GVD medium by introducing normal-GVD in free space [Fig.~\ref{Fig:DataReversal}(a)], and similarly enhance the dispersion in the anomalous-GVD regime by introducing anomalous GVD in free space [Fig.~\ref{Fig:DataReversal}(d)]. Comparing Fig.~\ref{Fig:DataReversal}(a,d) to Fig.~\ref{Fig:DataProfiles}(b,c) confirms the enhanced pulse broadening. Second, the GVD experienced by the wave packet in the medium can be \textit{cancelled} $k_{2\mathrm{m}}'\!=\!0$ by setting $k_{2\mathrm{a}}\!=\!-n_{\mathrm{m}}k_{2\mathrm{m}}-\tfrac{\Delta}{c\omega_{\mathrm{o}}}$ [Fig.~\ref{Fig:DataReversal}(b,e)], which is the scenario dealt with above in Fig.~\ref{Fig:DataProfilesNormalMaterialGVD}.

Third, the effective GVD experienced by the wave packet in the medium can be \textit{inverted} as shown in Fig.~\ref{Fig:DataReversal}(c,f). By GVD-inversion we mean that the effective GVD coefficient in the medium $k_{2\mathrm{m}}'$ has the opposite sign as that of the intrinsic chromatic dispersion in the medium $k_{2\mathrm{m}}$ (of course, the magnitudes need not be equal). In Fig.~\ref{Fig:DataReversal}(c), the ST wave packet traveling in ZnSe in the normal-GVD regime instead encounters \textit{anomalous} GVD. Here the anomalous GVD introduced in free space overcomes the normal GVD in the medium and renders it effectively an anomalous-GVD medium. Similarly, the ST wave packet in Fig.~\ref{Fig:DataReversal}(f) traveling in the anomalous-GVD medium encounters instead normal GVD. This can be useful in exploiting media that have desirable nonlinear coefficients for particular interactions but whose sign of GVD at the wavelength of interest is opposite of what is needed.

\section{Discussion and conclusion}

We emphasize again the distinction between `dispersion compensation' and `dispersion cancellation'. In the former, \textit{after} a conventional pulse traverses a dispersive medium, an optical system compensates for the dispersion (in principle of any order or sign) encountered by removing the accumulated spectral phase. This can be accomplished using a $4f$ spectral phase modulator \cite{Weiner00RSI} or other systems. By `dispersion cancellation' we refer to modifying the structure of the optical field by introducing angular dispersion, such that it propagates invariantly in the dispersive medium. Whereas conventional angular dispersion (in TPFs) can cancel normal GVD but \textit{not} anomalous GVD, we have demonstrated here that non-differentiable angular dispersion (in ST wave packets) enables GVD-cancellation in both the normal and anomalous regimes.

Although the existence of propagation-invariant ST wave packets in presence of normal or anomalous GVD was known theoretically, the lack of experimental strategies for producing non-differentiable angular dispersion precluded putting these predictions to test in the linear regime. Another obstacle faced previously is that the required ST wave packets were of the `baseband' class; i.e., their spatial spectra are centered at $k_{x}\!=\!0$ \cite{Yessenov19PRA}. Until recently, all experimentally generated ST wave packets in free space were of the `sideband' variety; i.e., there spatial spectra are centered at $k_{x}\!\neq\!0$ and the low spatial frequencies in the vicinity of $k_{x}\!=\!0$ are excluded on physical grounds \cite{Yessenov19PRA}. Examples include focus-wave modes \cite{Brittingham83JAP,Reivelt00JOSAA,Reivelt02PRE} and X-waves \cite{Lu92IEEEa,Saari97PRL}. Both of these obstacles are overcome by exploiting the universal angular dispersion synthesizer in \cite{Hall21OEUniversal}. Although such an approach introduces arbitrary angular dispersion in one transverse dimension only, recent progress has extended this strategy to both transverse dimensions \cite{Guo21Light,Pang2021OL,Yessenov2022Localized3D}.

Although a previous experiment demonstrated normal-GVD cancellation in silica using modified X-waves \cite{Sonajalg96OL,Sonajalg97OL}, no attempts at cancelling anomalous GVD by exploiting focus-wave modes, X-waves, or other sideband ST wave packets have been reported. Baseband ST wave packets have been synthesized via energy-inefficient spatio-temporal amplitude filtering for cancelling anomalous \cite{Dallaire09OE} and normal \cite{Jedrkiewicz13OE} GVD, but propagation invariance in presence of dispersion was not verified. Finally, theoretical studies have uncovered a host of structural field transitions for dispersion-free ST wave packets in dispersive media that have no analogs in free space, including a transition from X-shaped to O-shaped profiles while tuning the group velocity in presence of anomalous GVD \cite{Malaguti08OL}, and even more complex transitions in the normal-GVD regime \cite{Malaguti09PRA}. All such transitions occur at a \textit{fixed} wavelength (in contrast to \cite{Porras05OL,Panov21PRA} where the transition requires changing the GVD sign). None of these phenomena have been observed to date, and an O-shaped ST wave packet has not yet been reported. We anticipate that the work presented here can provide the platform for studying these structural dynamics in dispersive media.

In conclusion, we have realized -- for the first time to the best of our knowledge -- dispersion-free propagation in dispersive media symmetrically in the normal- \textit{and} anomalous-GVD regimes. By incorporating non-differentiable angular dispersion into a pulsed field we produce ST wave packets whose group velocity and GVD coefficient can be tuned in free space independently of each other. We have confirmed dispersion-free propagation of 200-fs pulses at a wavelength $\lambda_{\mathrm{o}}\!\approx\!1$~$\mu$m in ZnSe (normal GVD) and chirped Bragg mirrors (anomalous GVD). Moreover, because the GVD in the medium combines additively with the GVD introduced into the ST wave packet in free space, we have succeeded in demonstrating GVD-inversion: the wave packet experiences normal GVD while propagating in a medium in its anomalous-GVD regime, and vice versa. Moreover, we have demonstrated this unprecedented level of GVD control independently of the wave-packet group velocity, which can be tuned separately. These results are useful in multi-wavelength nonlinear interactions and quantum optics in long crystals. 

\section*{Appendix: Coupling a tilted-pulse front to a dispersive medium}

In an on-axis TPF in free space, the propagation angle with respect to the $z$-axis takes the general form $\varphi_{\mathrm{a}}(\omega_{\mathrm{o}}+\Omega)\!\approx\!\varphi_{\mathrm{a}}^{(1)}\Omega+\tfrac{1}{2}\varphi_{\mathrm{a}}^{(2)}\Omega^{2}$, with $\varphi(\omega_{\mathrm{o}})\!=\!0$, $\varphi_{\mathrm{a}}^{(n)}\!=\!\tfrac{d\varphi_{\mathrm{a}}}{d\omega}\big|_{\omega_{\mathrm{o}}}$, and we similarly expand $k_{x}$ and $k_{z}$, $k_{x}(\omega)\!\approx\!k_{x}^{(0)}+k_{x}^{1)}\Omega+\tfrac{1}{2}k_{x}^{(2)}\Omega^{2}$, and $k_{z}(\omega)\!\approx\!k_{z}^{(0)}+k_{z}^{(1)}\Omega+\tfrac{1}{2}k_{z}^{(2)}\Omega^{2}$; where $k_{x}^{(0)}\!=\!0$, $ck_{x}^{(1)}\!=\!\omega_{\mathrm{o}}\varphi_{\mathrm{a}}^{(1)}$, $c\omega_{\mathrm{o}}k_{x}^{(2)}\!=\!2\omega_{\mathrm{o}}\varphi_{\mathrm{a}}^{(1)}+\omega_{\mathrm{o}}^{2}\varphi_{\mathrm{a}}^{(2)}$, $k_{z}^{(0)}\!=\!k_{\mathrm{o}}$, $ck_{z}^{(1)}\!=\!1$, and $c\omega_{\mathrm{o}}k_{z}^{(2)}\!=\!-(\omega_{\mathrm{o}}\varphi_{\mathrm{a}}^{(1)})^{2}$ \cite{Porras03PRE2}. The last equation indicates that only anomalous GVD can be produced in free space on-axis with conventional angular dispersion. In a dispersive medium, the expansion coefficients for $k_{x}$ and $k_{z}$ are: $k_{x}^{(0)}\!=\!0$, $ck_{x}^{(1)}\!=\!n_{\mathrm{m}}\omega_{\mathrm{o}}\varphi_{\mathrm{m}}^{(1)}$, $c\omega_{\mathrm{o}}k_{x}^{(2)}\!=\!2\widetilde{n}_{\mathrm{m}}\omega_{\mathrm{o}}\varphi_{\mathrm{m}}^{(1)}+n_{\mathrm{m}}\omega_{\mathrm{o}}^{2}\varphi_{\mathrm{m}}^{(2)}$, $k_{z}^{(0)}\!=\!n_{\mathrm{m}}k_{\mathrm{o}}$, $ck_{z}^{(1)}\!=\!\widetilde{n}_{\mathrm{m}}$, and $c\omega_{\mathrm{o}}k_{z}^{(2)}\!=\!c\omega_{\mathrm{o}}k_{2}-n_{\mathrm{m}}(\omega_{\mathrm{o}}\varphi_{\mathrm{m}}^{(1)})^{2}$.

Because $k_{x}$ is invariant across a planar interface at normal incidence, matching the first-order expansion coefficients for $k_{x}$ yields $\varphi_{\mathrm{a}}^{(1)}\!=\!n_{\mathrm{m}}\varphi_{\mathrm{m}}^{(1)}$, which can be recognized as the law of refraction for TPFs at normal incidence. Dispersion-free propagation in the medium $k_{z}^{(2)}\!=\!0$ requires that $\omega_{\mathrm{o}}\varphi_{\mathrm{a}}^{(1)}\!=\!\tan{\delta_{\mathrm{a}}^{(1)}}\!=\!\sqrt{n_{\mathrm{m}}c\omega_{\mathrm{o}}k_{2\mathrm{m}}}$. Only normal GVD $k_{2\mathrm{m}}\!>\!0$ can be cancelled. The corresponding TPF in free space has a GVD coefficient $k_{z}^{(2)}\!=\!-n_{\mathrm{m}}k_{2\mathrm{m}}$. Therefore, GVD-cancellation requires exercising control over only first-order angular dispersion. To simplify the synthesis of the TPF in free space, we set $\varphi_{\mathrm{a}}^{(n)}\!=\!0$ for $n\!\geq\!2$. This assumption does \textit{not} lead to the elimination of $\varphi_{\mathrm{m}}^{(2)}$, which is given by $
n_{\mathrm{m}}\omega_{\mathrm{o}}^{2}\varphi_{\mathrm{m}}^{(2)}\!=\!2(1-\tfrac{\widetilde{n}_{\mathrm{m}}}{n_{\mathrm{m}}})\omega_{\mathrm{o}}\varphi_{\mathrm{a}}^{(1)}$. The transverse wave number is $k_{x}(\omega)\!=\!\tfrac{\Omega}{\omega_{\mathrm{o}}}\tfrac{\omega}{c}\tan{\delta_{\mathrm{a}}^{(1)}}$, which is differentiable with respect to $\omega$ everywhere. In free space, $k_{x}(\omega)\!=\!\tfrac{\omega}{c}\sin\{\varphi_{\mathrm{a}}(\omega)\}$, so that $\varphi_{\mathrm{a}}(\omega)\!\approx\!\tfrac{\Omega}{\omega_{\mathrm{o}}}\tan{\delta_{\mathrm{a}}^{(1)}}$.

Although the TPF in the material is GVD-free, higher-order dispersion terms nevertheless exist because of the $\varphi_{\mathrm{m}}^{(2)}$ term,. Of course one may eliminate the $\varphi_{\mathrm{m}}^{(2)}$ term by including an appropriate $\varphi_{\mathrm{a}}^{(2)}$ term in free space. However, this would add to the complexity of the system. Indeed, no known optical device -- besides the universal angular-dispersion synthesizer \cite{Hall21OEUniversal} -- has reported independent control over both $\varphi_{\mathrm{a}}^{(1)}$ and $\varphi_{\mathrm{a}}^{(2)}$. 

\section*{ACKNOWLEDGMENTS}
The authors acknowledge the support of the Office of
Naval Research (ONR, Grant Nos. N00014-17-1-2458 and
N00014-20-1-2789).

The authors declare no conflicts of interest.

\section*{DATA AVAILABILITY}
The data that support the findings of this study are available from the corresponding author upon reasonable request.

\bibliography{diffraction}

\end{document}